\documentclass[
aps,
prl,
twocolumn
]{revtex4-2}

\usepackage{graphicx}
\usepackage{dcolumn}
\usepackage{bm}
\usepackage{physics}
\usepackage{float}
\usepackage{amssymb}
\usepackage{amsmath}
\usepackage[unicode]{hyperref}
\usepackage{subcaption}
\usepackage[font=small,
   justification=Justified,
   format=plain]{caption}

\DeclareMathOperator{\sgn}{sgn}

\begin{document}

\title{Beyond spin-charge separation: Helical modes and topological quantum phase transitions in one-dimensional Fermi gases with spin-orbit and Rabi couplings}

\author{Xiaoyong Zhang} \thanks{xzhang840@gatech.edu}
\author{C. A. R. S\'a de Melo} 
\affiliation{School of Physics, Georgia Institute of Technology, Atlanta, GA 30332, USA}

\date{\today}

\begin{abstract}
Motivated by the  experimental observation of spin-charge separation in one-dimensional interacting Fermi gases, we investigate these systems in the presence of spin-orbit coupling and Rabi fields. We demonstrate that 
spin-charge-separated modes evolve into helical collective modes due to the special mixing of spin and charge induced by spin-orbit coupling and Rabi fields.
We obtain the phase diagram of
chemical potential versus Rabi fields for given spin-orbit coupling and interactions, and find several topological quantum phase transitions of the Lifshitz type. We show that the velocities of the collective modes are nonanalytic at the boundaries between different phases. Lastly, we analyze the 
charge-charge, spin-charge and spin-spin dynamical structure factors to show that the dispersions, spectral weights and helicities of the collective modes
can be experimentally extracted
in systems such as 
$^{6}{\rm Li}$, $^{40}{\rm K}$ and $^{173}{\rm Yb}$.
\end{abstract}

\maketitle

\textit{Introduction:} 
While spin and charge are intrinsic properties of elementary particles, in one dimension (1D), interactions are responsible for the separation of spin and charge leading to spin-density (SDW) and charge-density (CDW) waves that propagate with different velocities. Spin-charge separation is theoretically described by the Tomonaga-Luttinger liquid model \cite{1950-tomonaga,1963-luttinger,1981-haldane, 2003-giamarchi,2023-altland,
2012-glazman} 
in condensed matter physics (CMP), but is regarded as a general phenomenon of a large variety of quantum fields: non-Abelian Yang-Mills theory describing knotted strings as stable solitons \cite{2005-walet,2007-niemi-b,2007-niemi}, supersymmetric gauge theory characterizing magnetic superconductors \cite{1998-mavromatos}, and quark-lepton unification theory suggesting the similarity between spinons and neutrinos \cite{2015-xiong}.
A few experiments in condensed matter have claimed observing spin-charge separation: angle-resolved photoemission in $\text{SrCuO}_2$ \cite{1996-maekawa,2006-kim} and tunneling spectroscopy in $\text{GaAs/AlGaAs}$ heterostructures \cite{2002-west,2005-west} at low temperatures.
Very recently, spin-charge separation was also observed in ultracold gases ($^{6}\text{Li}$) as a function of interactions and temperature~\cite{2022-hulet,2023-hulet}. A major experimental advantage of ultracold gases over condensed matter and high energy systems is the tunability of interactions, temperature, density and external fields, permitting for a thorough exploration of spin-charge separation and mixing \cite{2022-hulet, 2023-hulet}. This tunability allows for investigations of the interplay between spin and charge degrees of freedom in unprecedented ways like, for instance, as a function of synthetic spin-orbit coupling, Rabi (spin-flip) fields, density or chemical potential.

In CMP, spin-orbit coupling (SOC) is a relativistic quantum mechanical effect that entangles the spin of a particle to its spatial degrees of freedom. SOC plays an important role in spin-Hall systems ~\cite{2015-jungwirth,2021-hayashi} and topological insulators~\cite{2010-kane,2011-zhang}
and superconductors~\cite{2017-ando,2022-li,2023-hu}. 
However, tunability of SOC, spin-flip fields, density or chemical potential is limited. There are two common types of SOC: the Rashba~\cite{rashba1961combined} and the Dresselhaus~\cite{1955-dresselhaus} terms, that have been discussed in the context of semiconductors~\cite{1984-rashba,2008-macdonald,2017-sirringhaus,2021-yang,2017-culcer,2021-lau}. 
In ultracold atoms, SOC is synthetically created using two-photon Raman transitions instead of originating from relativistic effects~\cite{2011-spielman-b,2013-spielman}. 
This makes SOC tunable in bosonic \cite{2011-spielman,2014-Bloch,2022-tarruell} and fermionic \cite{2012-zwierlein,2015-fallani,2017-ye} quantum gases, where equal mixtures of Rashba and Dresselhaus (ERD) terms have been created~\cite{2011-spielman},
as well as, 
Rashba-only (RO)~\cite{2016-spielman,2016-zhang,2021-spielman}. Furthermore, 
Dresselhaus-only (DO) and arbitrary mixtures of Rashba and Dresselhaus terms have also been suggested~\cite{2012-sademelo}.

Inspired by the observation of spin-charge separation in 1D interacting Fermi gases~\cite{2022-hulet}, we explore the effects of SOC and Rabi fields on spin-charge separation and show that spin and charge density modes evolve into spin-charge-mixed helical collective modes.
We investigate the phase diagram of
chemical potential versus Rabi fields for given SOC and interactions, revealing several topological quantum phase transitions of the Lifshitz type. We demonstrate that the velocities of the helical collective modes are nonanalytic at the boundaries between different phases, and show that the 
charge-charge, spin-charge and spin-spin dynamical structure factors~\cite{2022-hulet, 2023-hulet, 2022-moritz-a, 2022-moritz-b} 
reveal the dispersions, spectral weights
and helicities of the collective modes in experimentally accessible systems such as 
$^{6}{\rm Li}$, $^{40}{\rm K}$ and $^{173}{\rm Yb}$.

\textit{Hamiltonian:}
To describe the rich SOC physics outlined above, we consider the experimentally realized
kinetic energy operator~\cite{2011-spielman}
\begin{equation}
\label{eqn:hamiltonian-soc}
{\hat K} = 
\epsilon_k I-h_x\sigma_x-h_z (k)\sigma_z,
\end{equation}
where $k$ is the momentum along the $x$ direction, 
$\epsilon_k = {(k^2+k_T^2)}/{2m}$ is the shifted kinetic energy, $k_T$ is the spin-dependent momentum shift, $h_x$ is the Rabi field, and $h_z (k) ={k k_T}/{m}$ is the SOC.
We diagonalize ${\hat K}$  
using a momentum-dependent ${\rm SU(2)}$ rotation $U(k)$
to obtain the eigenvalues 
$E_A (k) = \epsilon_k-h_\text{eff} (k)$ and $E_B (k) =\epsilon_k + h_\text{eff} (k)$, where $h_\text{eff} (k) =\sqrt{h_x^2+h_z^2 (k) }$.  

\begin{figure}[ht]
\centering
\includegraphics[width=7.0 cm]{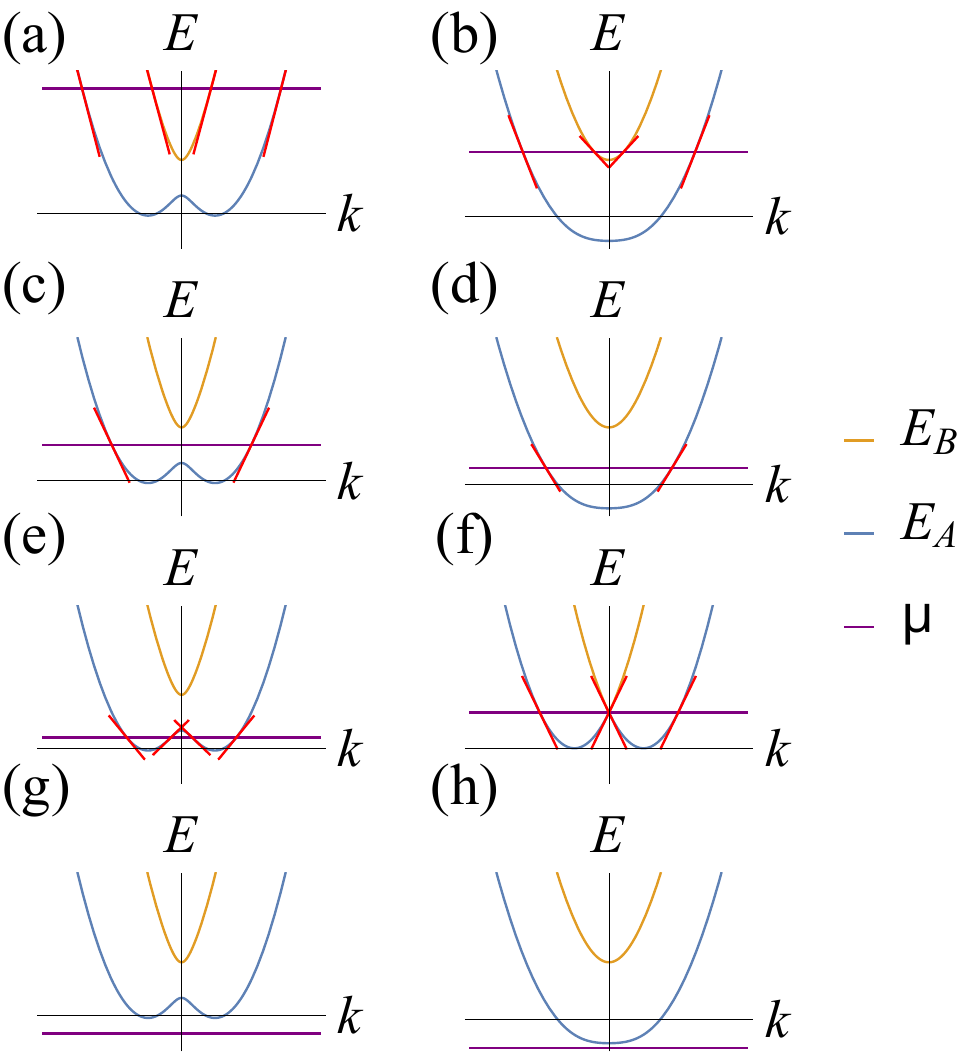}
\caption{Schematic plots of $E_{A} (k)$ (blue solid line) and $E_{B} (k)$ (orange solid line) showing effects of SOC $(k_T)$ and Rabi fields $(h_x)$. The red solid lines are linearizations around  Fermi points, and the horizontal purple solid lines indicate 
$\mu$. In panels (a), (c), (e) and (g), $E_A (k)$ has double minima, since $\vert h_x \vert/ E_T < 2$. In panels (b), (d), and (h), $E_{A} (k)$ has a single minimum, 
since $\vert h_x\vert/E_T \ge 2$. In panel (f), $h_x = 0$, and the system is equivalent to no SOC due to the spin-gauge symmetry.
}
\label{fig:1}
\end{figure}

In Fig.~\ref{fig:1}, $E_A (k)$ and $E_B (k)$ are shown for various situations. 
When $h_x\neq 0$, $E_A (k)$ and $E_B (k)$ are non-degenerate with $E_A (k)$ having either double minima $(\vert h_x \vert < 2 E_T)$, shown in Fig. \ref{fig:1}(a,c,e,g), or a single minimum 
$(\vert h_x \vert \ge 2 E_T)$
seen in Fig. \ref{fig:1}(b,d,h), where 
$E_T = k_T^2/2m$. 
When $h_x=0$, we have $h_\text{eff} (k) = \vert h_z (k) \vert$ and the two bands intersect at $k=0$, as shown in Fig. \ref{fig:1}(f). The creation operators for the energy basis
are $(a_{k,A}^\dagger\ a_{k,B}^\dagger)=(c_{k,\uparrow}^\dagger\ c_{k,\downarrow}^\dagger)U^\dagger(k)$, where
$c^\dagger_{k, s}$ are the creation operators with momentum $k$ and spin $s = \{\uparrow, \downarrow\}$. The operators $a^\dagger_{\alpha}$ create helical fermions, because the matrix $U(k)$ has an 
${\rm SU(2)}$ momentum-dependent rotation angle 
$\theta (k_T, h_x, k)$, 
when both $k_T$ and $h_x$ are non-zero. 

Interactions for two-internal-state $^6{\rm Li}$, $^{40}{\rm K}$ or $^{173}{\rm Yb}$ are 
$
{\hat V}=V_0 \int_0^L\dd x^\prime 
\int_0^L\dd x  {\hat n}_{\uparrow}(x){\hat n}_{\downarrow}(x^\prime) f(\vert x-x^\prime \vert),
$
where ${\hat n}_s(x)=\psi_s^\dagger(x)\psi_s(x)$ is the local density operator
with $s= \{\uparrow, \downarrow\}$ and $L$ is the length of system.
The parameter $V_0$, with dimensions of 
energy, controls the strength and the dimensionless function $f(\vert x- x^{\prime}\vert )$ controls the range $r_0$ of the interaction. In momentum space
\begin{equation}
\label{eqn:hamiltonian-interaction-momentum-space}
{\hat V}=\frac{V_0}{L}\sum_q  {\hat \rho}_{\uparrow} (q) {\hat \rho}_{\downarrow} (-q){\tilde f}(q),
\end{equation}
where ${\hat \rho}_{s} (q)=\sum_{k}c^\dagger_{k+q,s}c_{k,s}$ is the Fourier transform of the local density operator ${\hat n}_s(x)$, and ${\tilde f}(q) = 
\int d \eta e^{-{\rm i}q\eta} f(\vert \eta \vert)$ is the Fourier transform of $f(\vert x-x^{\prime} \vert)$
with dimensions of length.
When $h_x=0$, the spin-gauge transformation $\psi_\uparrow(x)\rightarrow e^{{\rm i}k_T x}\psi_\uparrow(x),\ \psi_\downarrow(x)\rightarrow e^{-{\rm i} k_T x}\psi_\downarrow(x)$ 
gauges away $k_T$ in the kinetic energy without changing the interaction ${\hat V}$ (spin-gauge symmetry). 

\textit{Bosonization:}
To describe the low-energy excitations, we use the bosonization technique.  
We approximate ${\hat K}$ in Eq.~(\ref{eqn:hamiltonian-soc}) by linearizing it around the chemical potential $\mu$. 
There are four general cases shown in Fig.~\ref{fig:1}: (I) 
$\mu$ intersects twice $E_A (k)$ and twice $E_B (k)$, see Figs.~\ref{fig:1}(a,b); (II) 
$\mu$ intersects only band $E_A (k)$ twice, see Figs.~\ref{fig:1}(c,d); (III) $\mu$ intersects $E_A (k)$ four times, see 
Fig.~\ref{fig:1}(e); 
(IV) $\mu$ does not intersect $E_A (k)$ or $E_B (k)$, see Fig.~\ref{fig:1}(g,h). Intersection of $\mu$ to either 
$E_A (k)$ or $E_B (k)$ allows for linearization of dispersions, shown as red lines in Fig.~\ref{fig:1}. We 
label particles by indices 
$\{r, \alpha\}$: $r$ describes left ($L$) or right ($R$) moving fermions; $\alpha$ labels $A$ or $B$ indicating linearization of 
$E_A (k)$ or 
$E_B (k)$. In the special case of Fig.~\ref{fig:1}(e), we set 
$\alpha = A^{(1)}$ for the outer red lines and 
$\alpha = A^{(2)}$ for the inner red lines, since only $E_A (k)$ is crossed by $\mu$.  The linearized kinetic energy operator is
\begin{equation}
\label{eqn:hamiltonian-kinetic-linearized}
{\hat K} = \sum_{k,r,\alpha} u_{\alpha}\big[\sgn(r)k-k_{\mu,\alpha}\big] a^{\dagger (r)}_{k,\alpha}a^{(r)}_{k,\alpha},
\end{equation}
where $u_{\alpha}$ is the Fermi velocity $d E_{\alpha}(k)/dk$ at $k = k_{\mu,\alpha}$, with $k_{\mu,\alpha}$ being 
the positive momentum where 
$\mu$ intersects the relevant band. 
Both $u_{\alpha}$ and $k_{\mu,\alpha}$ 
depend on $k_T$, $h_x$ and $\mu$.
The function $\sgn(r)$ refers to the sign of $r$, where $\sgn (R) = +1$ and $\sgn(L) = -1$.

We bosonize ${\hat K}$ in Eq.~(\ref{eqn:hamiltonian-kinetic-linearized}) using
the transformation
\begin{equation}
\label{eqn:boson-operator}
b_{q,\alpha}^\dagger=\sqrt{\frac{2\pi}{\vert q\vert L}}
\sum_{r}
\Theta\big(-\sgn(r)q\big){\hat \rho}_{\alpha}^{\dagger (r)}(q),
\end{equation}
where 
${\hat \rho}_{\alpha}^{(r)} (q)$ are the fermion density 
operators, and 
$\Theta(x)$ is the
step function.
This leads to 
$
{\hat K}
=
\sum_{q,\alpha} \vert q  u_\alpha \vert b_{q,\alpha}^\dagger b_{q,\alpha}
$
for the kinetic energy and to
\begin{equation}
\label{eqn:hamiltonian-interaction-eigenbasis}
{\hat V}=
\frac{V_0}{\pi} 
\sum_{q>0}
q {\boldsymbol \Phi}^\dagger (q) {\bf g}{\boldsymbol \Phi} (q)
\end{equation}
for the interaction, where ${\boldsymbol \Phi}(q)$ is either a two-component vector or a four-component vector, and ${\bf g}$ is either a $2\times 2$
or a $4\times 4$ matrix. 
In case (I), where bands $A$ and $B$ are
crossed,
${\boldsymbol \Phi}^T(q)=\begin{pmatrix}
b_{q,A}^\dagger&
b_{q,B}^\dagger&
b_{-q,A}&
b_{-q,B}
\end{pmatrix}.
$
In case (II), where 
band $A$ is crossed 
at two Fermi points,
${\boldsymbol \Phi}^T (q)=\begin{pmatrix}
b_{q,A}^\dagger&
b_{-q,A}
\end{pmatrix}.
$
In case (III), where band $A$ is crossed
at four Fermi points,
${\boldsymbol \Phi}^T (q)=\begin{pmatrix}
b_{q,A^{(1})}^\dagger&
b_{q,A^{(2)}}^\dagger&
b_{-q,A^{(1)}}&
b_{-q,A^{(2)}}
\end{pmatrix}.
$
The matrix elements of ${\bf g}$ are $g_{ij} = \eta_{ij}{\tilde f} (q \to 0)$, 
having dimensions of length, while 
$\eta_{ij}$ is dimensionless and depends on 
$k_T$, $h_x$ and $\mu$.

\begin{figure}[ht]
\centering
\includegraphics[width= 7.0 cm]{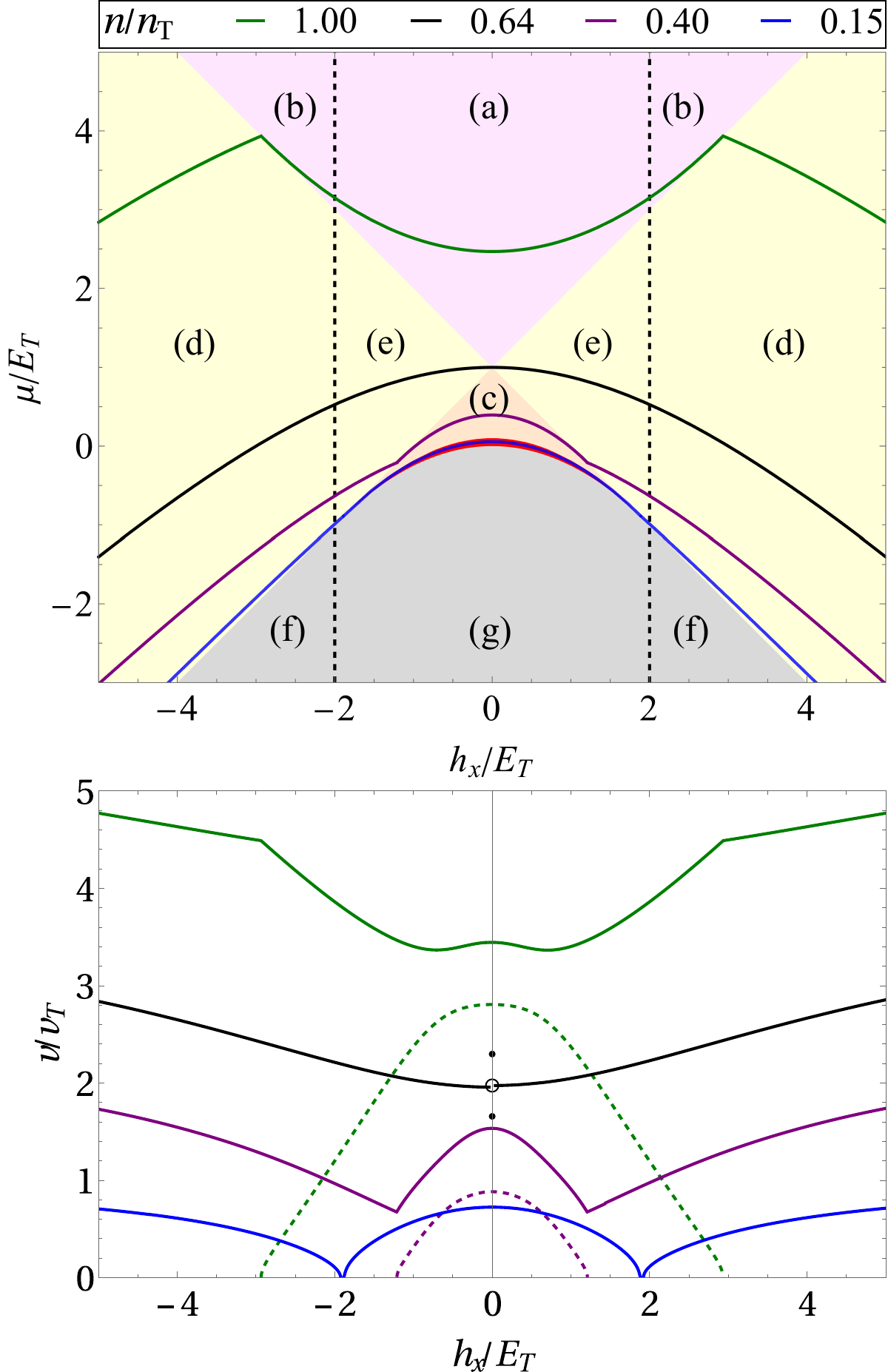}
\caption{Phase diagram and velocities of collective modes. In panel (a), we show the phase diagram of $\mu/E_T$ versus $h_x/E_T$, where $E_T=k_T^2/2m$ for $V_0/E_T = 2$. The pink, yellow, orange and gray regions represent phases with four (two), two (one), four (two) and zero (zero) Fermi points (collective modes), respectively separated by topological quantum phase (Lifshitz) transitions. In the red region the lower velocity mode is unstable. The vertical black dashed lines at $\vert h_x \vert/E_T = 2$ separate regions where $E_{A} (k)$ has single minimum $\vert h_x \vert/E_T < 2$ or double minima $\vert h_x \vert/E_T > 2$. The letters $a, b, c, e, f, g$ represent the dispersions shown in Fig.~\ref{fig:1}. The green, black, purple and blue solid lines represent fixed densities $n/n_T = \{ 1.00, 0.64, 0.40, 0.15 \}$, respectively.
In panel (b), we show the velocities $v/v_T$ of collective modes versus $h_x/E_T$ in green (solid and dashed) for $n_/n_T = 1.00$,
in black solid for $n/n_T = 2/\pi \approx 0.64$, in purple (solid and dashed) 
for $n_/n_T = 0.40$ and
in blue (solid) for $n/n_T = 0.15$. The velocities are nonanalytic at the boundaries of phases, and the line $h_x/E_T$ represents the spin-charge separation locus.
}
\label{fig:2}
\end{figure}

\textit{Collective modes:}
The bosonized Hamiltonian ${\hat H} = 
{\hat K} + {\hat V}$ is diagonalized  via a Bogoliubov transformation ${\bf B}$ leading to 
boson operators 
$
\Psi^T(q)=
\begin{pmatrix}
d_{q,1}^\dagger&  
d_{q,2}^\dagger&
d_{-q,1}&
d_{-q,2}
\end{pmatrix}
$ 
for cases (I) and (III), where the number of collective modes is $N_C = 2$ and to 
$
\Psi^T(q)=
\begin{pmatrix}
d_{q,1}^\dagger&  
d_{-q,1}
\end{pmatrix}
$ 
for case (II), where
$N_C = 1$.
Barring any instabilities, $N_C =  N_F/2$, where $N_F$ is the number of Fermi points at $\mu$.
The transformation
$\Psi(q)= 
{\bf B}\Phi (q)$
leads to the diagonalized Hamiltonian
\begin{equation}
\label{eqn:hamiltonian-diagonal-form}
{\hat H}=\sum_{q,m}\vert q\vert v_m d_{q,m}^\dagger d_{q,m}
+ \omega_{\Omega},    
\end{equation}
where $v_m \ge 0$ is the velocity of collective mode $m$ that depends on $\{k_T, h_x, \mu\}$ for given $V_0$. The ground state energy is
$\omega_{\Omega} = E_\Omega - K_\Omega$,
where $E_{\Omega} = \sum_{q > 0,m} \vert q v_{m} \vert$ and 
$K_{\Omega} = \sum_{q > 0,\alpha}
\vert q u_{\alpha}\vert$.

In Fig.~\ref{fig:2}(a), we show the phase diagram of $\mu/E_T$ versus $h_x/k_T$,
for $V_0 /E_T = 2$.
In the pink region,
$N_F = 4$ as in Fig.~\ref{fig:1}(a)-(b) and
$N_C = 2$.
 In the yellow region, $N_F = 2$ as in Fig~\ref{fig:1}(c)-(d) and
 $N_C = 1$. In the orange region, 
 $N_F = 4$ as in Fig.~\ref{fig:1}(e) and 
 $N_C = 2$. For the vertex separating the pink, yellow and orange regions  
 ($h_x/E_T = 0$ and $\mu/E_T = 1$), 
 $N_F = 4$ as in 
 Fig.~\ref{fig:1}(f) and $N_C =2$; at this location there is perfect spin-charge separation.
 In fact, along the line $h_x/E_T = 0$, there is spin-charge separation
 also in the pink and orange regions, due to the spin-gauge symmetry.
 In the red region, the lower-velocity collective mode becomes unstable. This instability arises from the breakdown of
 the positive definiteness of the Hamiltonian with respect to the complex symplectic group ${\rm Sp}(4, {\bf C})$
 that the Bogoliubov transformation satisfies~\cite{1936-williamson}. 
 In the gray region, $N_F = 0$ and $N_C = 0$, since $\mu$ lies below the energy bands as in Fig.~\ref{fig:1}(g)-(h); this the zero density limit. 
 
 Furthermore, in Fig.~\ref{fig:2}(a), the lines or points separating different phases indicate topological quantum phase transitions of the Lifshitz-type~\cite{1960-lifshitz}, where the Fermi {\it surfaces} change from four to two to zero points, depending on $\mu/E_T$ and 
 $h_x/E_T$. The vertical black dashed lines at 
 $h_x/E_T = \pm 2$ separate regions where the lower band has two minima
 $(\vert h_x \vert/E_T < 2)$ from regions where 
 the lower band has one minimum 
 $(\vert h_x \vert/E_T > 2)$
 and the line at
 $h_x/E_T = 0$ represents the locus of spin-charge separation. Since experiments are performed at fixed densities,
 the solid lines are constant density curves in units of $n_T = k_T$. 
 The densities are as follows: $n/n_T = 1.00$ (green line),  
 $n/n_T = 2/\pi \approx 0.64$ (black line), 
 $n/n_T = 0.40$ (purple line),
 $n/n_T = 0.15$ (red line).

In Fig.~\ref{fig:2}(b), we show the collective mode velocities $v/v_T$ versus $h_x/E_T$, where $v_T = k_T/m$, at
$V_0/E_T = 2$
for fixed densities $n/n_T = 1.00$ (green solid and
dashes lines), $n/n_T = 0.64$ (black solid and dashed lines), $n/n_T = 0.40$
(purple solid and dashed lines), and
$n/n_T = 0.15$ (blue solid line). Notice that $v/v_T$ is continuous, but nonanalytic every time a phase boundary is crossed; this behavior is a direct consequence of a topological (Lifshitz) quantum phase transition. There are always two modes in the pink region, one mode in the yellow region, two modes in the orange region, one mode in the red region,
and no modes in the gray region.
For $h_x/E_T = 0$ there is perfect spin-charge separation, and when two modes exist the higher-velocity mode is associated with charge and the lower-velocity mode is associated with spin. 
As the density is lowered into the red region of Fig.~\ref{fig:2}(a), the spin mode becomes unstable, and only the charge mode survives.
In the absence of SOC 
$(k_T = 0)$ with $h_x \ne 0$, the collective modes are generally a mix of charge and spin density waves,  
but they are nonhelical. However,  
when both $k_T \ne 0$ and 
$h_x \ne 0$, the collective modes are a mix of charge and spin density waves with a helical structure, which is carried over by the momentum-dependent rotation angles $\theta (k_T, h_x, k)$ of the ${\rm SU(2)}$ rotation matrices $U(k)$.

To visualize helical modulations,
we write 
$
d_{q, m} = 
\sum_r\big[ A_m^{(r)} (q) {\hat \rho}^{(r)} (q)   
+ {\bf B}_m^{(r)} (q) \cdot
{\hat {\boldsymbol \sigma}}^{(r)} (q)
\big]
$
in terms of the Fourier transforms 
${\hat \rho}^{(r)} (q)$
and ${\hat {\boldsymbol \sigma}}^{(r)} (q)$
of the charge-density
${\hat n}^{(r)} (x) = 
\sum_s
\psi_s^{\dagger (r)} (x) 
\psi_s^{(r)}(x)$
and 
spin-density 
${\hat {\bf S}}^{(r)} (x) = 
\sum_{s s^\prime}
\psi_s^{\dagger(r)} (x) 
(\boldsymbol \sigma)_{s s^{\prime}}
\psi_{s^\prime}^{(r)} (x)$ operators, respectively.
The collective mode operators in real space are 
$
{\cal D}_m (x)= 
\sum_{r} \int \dd x^{\prime} 
\big[
a_m^{(r)} (x - x^{\prime})
{\hat n}^{(r)} (x^{\prime}) 
+ 
{\bf b}_m^{(r)} 
(x - x^{\prime})
\cdot {\hat {\bf S}}^{(r)} (x^{\prime})
\big],  
$
where the spatial modulation and helicity of the vector fields
${\bf b}_{m}^{(r)} (x - x^{\prime})$ are controlled by 
$k_T$ and $h_x$. When $k_T$ and 
$h_x$ are both non-zero, all the modes present are helical, with $q > 0$ $(q < 0)$ modes having positive (negative) helicity. The global helicity of the Hamiltonian in Eq.~(\ref{eqn:hamiltonian-diagonal-form}) is zero, such that $q >0$
and $q < 0$ bosons for each mode are helical pairs.

For $^6{\rm Li}$, $^{40}{\rm K}$, and $^{173}\rm{Yb}$ typical SOC and Rabi fields are known~\cite{2012-zwierlein,2013-spielman-b, 2016-fallani}. Optical boxes in 1D have  typical dimensions from $10 \mu{\rm m}$ to $100\mu{\rm m}$~\cite{2021-hadzibabic}, and  number of atoms from a few~\cite{2021-lahaye} to thousands~\cite{2014-hadzibabic}.
As an example, we consider a 1D optical box with $L=50\mu{\rm m}$, a tight transverse confinement frequency
$
\omega_r=2\pi\times227.5\text{kHz}
$ 
~\cite{2022-hulet}, 
number of atoms $N=400$, interaction $V_0/E_T = 2$
in the zero-range limit $r_0 \to 0$.
For $^6{\rm Li}$, typical parameters are $k_T=2\pi\times (500 \text{nm})^{-1}$ for the SOC momentum transfer~\cite{2022-hulet}, $n/n_T=0.64$ for the density, and 
$a_{\rm 3D} \approx 917 a_0$ for 
the 3D scattering length, where $a_0$ is the Bohr radius. For these parameters, the  behavior for the collective mode velocities in $^{6}{\rm Li}$ corresponds to the $n/n_T = 0.64$ (black lines)
shown in Fig.~\ref{fig:2}(b).
For $^{40}{\rm K}$, $k_T=2\pi\times (768.86 \text{nm})^{-1}$~\cite{2013-spielman-b}, $n/n_T=0.98$, and 
$a_{\rm 3D} \approx 159 a_0$, and 
the behavior of the collective mode velocities is very
similar to that of $n/n_T=1$ (green lines), shown in Fig.~\ref{fig:2}(b).
For $^{173}{\rm Yb}$, typical parameters are $k_T=2\pi\times (289 \text{nm})^{-1}$~\cite{2016-fallani},  
$n/n_T=0.37$, and $a_{3D} \approx 91 a_0$, and the collective mode velocities behavior is just like the $n/n_T=0.40$ (purple lines) case, shown in Fig.~\ref{fig:2}(b).

\begin{figure}[ht]
\centering
\includegraphics[width=8.6 cm]{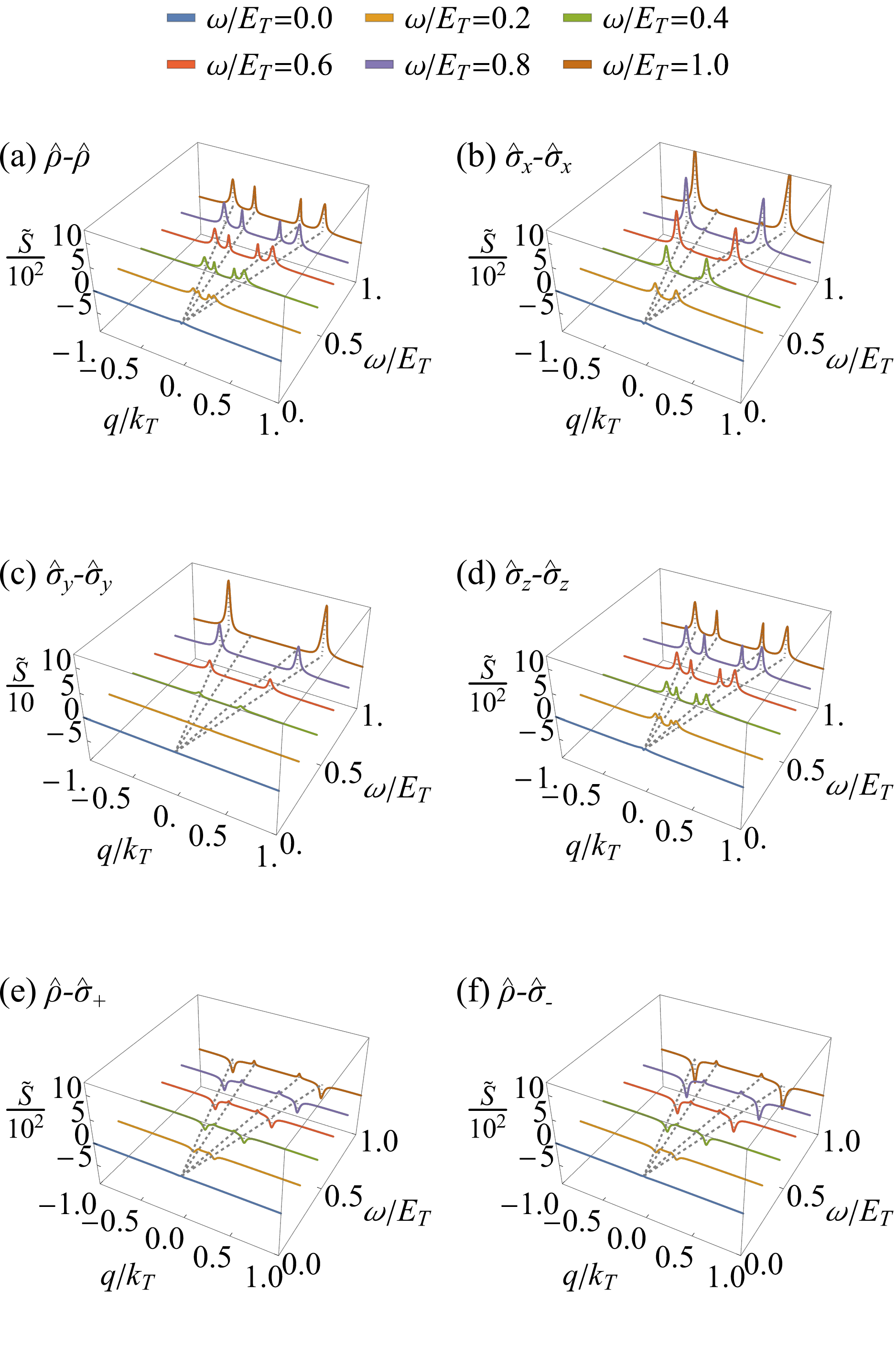}
\caption{Dimensionless DSF tensor 
${\widetilde S}_{ij} (q, \omega) = S_{ij} (q.\omega) E_T/k_T^2$, for 
$h_x/E_T=1.05$, $\mu/E_T =2.85$ $(n/n_T = 1.00)$, $V_0/E_T = 2$ $(a_{3D} \approx 159 a_0)$ and $k_T L = 408.60$ corresponding to 
typical values for $^{40}{\rm K}$, with energy broadening 
$\delta/E_T = 0.005$.
Panels (a) through (d) show the charge-charge 
and spin-spin responses: (a) ${\hat \rho}$-${\hat \rho}$; (b) ${\hat \sigma}_x$-${\hat \sigma}_x$; 
(c) ${\hat \sigma}_y$-${\hat \sigma}_y$;
and (d) ${\hat \sigma}_z$-${\hat \sigma}_z$.
Panels (e) and (f) display the charge-spin
${\hat \rho}$-${\hat \sigma}_+$ and 
${\hat \rho}$-${\hat \sigma}_-$ responses, respectively. The ${\hat \sigma}_y$-${\hat \sigma}_y$ response in (c) is weaker, hence it has a different scale. The gray dashed lines show the dispersing helical collective modes.
}
\label{fig:3}
\end{figure}

\textit{Response functions:}
Bragg scattering techniques have been used to measure velocities of charge and spin density collective modes~\cite{2022-moritz-a, 2022-moritz-b}, as well as to identify spin-charge separation in 
$^6{\rm Li}$~\cite{2022-hulet, 2023-hulet}. These experiments 
traditionally measure either charge or spin dynamical structure factors (DSF). Here, 
we go beyond that and investigate the spin-spin, charge-charge and spin-charge responses at $T = 0$
via the DSF tensor
$
{S}_{ij}(q,\omega) =
2\pi\sum_{m, p}\langle \Omega \vert {{\hat {\cal O}}_{i} (q)}\vert {m,p} \rangle \langle {m,p}\vert {{\hat {\cal O}}_{j} (-q)}
\vert \Omega \rangle 
\delta(\omega -\varepsilon_{p,m} + \omega_{\Omega}),
$
for the ground state $\vert \Omega \rangle $, in the K\"all\'en-Lehmann spectral representation~\cite{1952-kallen,1954-lehmann}. 
The operators ${\hat {\cal O}}_i (q)$, with $i = \{c, x, y, z\}$, are the charge $\hat {\cal O}_c (q) = {\hat \rho} (q)$ and
spin ${\cal O}_{\ell} (q) = {\hat \sigma}_{\ell} (q)$ operators, 
where $\ell = \{x, y, z\}$
or $\ell = \{+, -, z\}$, with 
$+$ $(-)$ labelling the spin raising
(lowering) operator ${\hat \sigma}_{+}$ 
$({\hat {\sigma}}_{-})$. 
The eigenstates of the bosonized Hamiltonian ${\hat H}$ given in Eq.~(\ref{eqn:hamiltonian-diagonal-form}) are $\vert m, p \rangle$ with 
corresponding eigenenergies $\varepsilon_{p, m}
= \vert p \vert v_{m}$, 
while $\omega_\Omega$ is the ground 
state energy. 

To calculate $S_{ij} (q, \omega)$, we implement 
the Moore-Penrose inverse~\cite{1920-moore,1955-penrose} 
and write the operators ${\hat {\cal O}}_{i}^{(r)} (q) $ in terms of the boson operators $d_{q,m}.$
When $q > 0$,  we obtain
$
{\hat {\cal O}}_{i}^{(r)}(q) =\sqrt{\frac{ \vert q\vert  L}{2\pi}}\sum_{m=1,2}\left(F_{im}^{(r)} (q) d_{-q,m}^\dagger+G_{im}^{(r)} (q) d_{q,m}\right), 
$
where the expression relating $q>0$ and $q<0$ is 
$
{\hat {\cal O}}_{i}^{(r)}(-q)
=
{\hat {\cal O}}_{i}^{\dagger (r)}(q)
$.
To obtain the matrix elements
$\langle \Omega \vert {{\hat {\cal O}}_{i} (q)}\vert {m,p} \rangle
$
and 
$
\langle {m,p}\vert {{\hat {\cal O}}_{j} (-q)}
\vert \Omega \rangle,
$
we use
${\hat {\cal O}}_{j}(q)={\hat {\cal O}}_{j}^{(R)}(q) + 
{\hat {\cal O}}_{j}^{(L)}(q)$
leading to
\begin{equation}
S_{ij} (q, \omega)
= 
\sum_m {\cal A}_{ij}^{(m)} (q)
\delta (\omega -\varepsilon_{q,m}
+ \omega_\Omega),
\end{equation}
where 
${\cal A}_{ij}^{(m)} (q)$
plays the role of the spectral weight tensor
with dimensions of
squared-density (squared-length in 1D), $\varepsilon_{q, m} = 
\vert q \vert v_m$ is the collective mode energy, and we
set $\omega_{\Omega} = 0$ to be our energy 
reference.
The Onsager reciprocal relation~\cite{1931-onsager-a,1931-onsager-b} for the DSF tensor is 
$
S_{ij}({q,\omega,k_T,h_x})=\epsilon_i\epsilon_jS_{ji}({-q,\omega,-k_T,-h_x}),
$
where $\epsilon_i$ is the parity of operators ${\hat {\cal O}}_i (x, t)$ under time-reversal.
For $i = c$ (charge density), $\epsilon_i = +1$, and for $i = \{x, y, z\}$ (spin density), $\epsilon_i = -1$.

In Fig.~\ref{fig:3}, we show matrix elements of 
${\widetilde S}_{ij} (q,\omega) = 
S_{ij} (q,\omega) E_T/k_T^2 = 
S_{ij} (q,\omega)/2m$
with energy broadening $\delta/E_T = 0.05$.
The parameters used are $h_x/E_T = 1.50$, $\mu/E_T= 2.85$
$(n/n_T = 1.00)$, $V_0/E_T = 2$
$(a_{3D} \approx 159 a_0)$, and 
$k_T L = 408.60$ corresponding to 
typical values for $^{40}{\rm K}$. 
These values correspond to a point in the pink region of Fig.~\ref{fig:2}(a), where there are two collective modes.

In all panels of Fig.~\ref{fig:3}, the gray dashed lines represent the two dispersing helical modes. In panels (a) through (d), we show the charge-charge and spin-spin responses:
(a) ${\hat \rho}$-${\hat \rho}$; (b) ${\hat \sigma}_x$-${\hat \sigma}_x$; 
(c) ${\hat \sigma}_y$-${\hat \sigma}_y$;
and (d) ${\hat \sigma}_z$-${\hat \sigma}_z$.
The ${\hat \sigma}_y$-${\hat \sigma}_y$ response in (c) is weaker, hence it has a different scale.
In panels (e) and (f), we show the charge-spin
${\hat \rho}$-${\hat \sigma}_+$ and 
${\hat \rho}$-${\hat \sigma}_-$ responses to reveal the helicity of the modes, 
since the responses are different.
We verified that for $h_x \ne 0$ and $k_T = 0$, the modes are non-helical, and that the only non-zero responses are ${\hat \rho}$-${\hat \rho}$, 
${\hat \sigma}_x$-${\hat \sigma}_x$; 
${\hat \rho}$-${\hat \sigma}_x$.
Furthermore, for $h_x = 0$ and any $k_T$, there is spin-charge separation, and the only non-zero responses are ${\hat \rho}$-${\hat \rho}$,
and ${\hat \sigma}_z$-${\hat \sigma}_z$.

\textit{Conclusions:}
We studied the phase diagram and collective modes of interacting one-dimensional Fermi systems with spin-orbit coupling and Rabi fields. We
indicated that topological quantum phase transitions of the Lifshitz type are induced
by spin-orbit coupling and Rabi fields. We found that the phase diagram exhibits regions with two, one or zero helical collective modes, which correlate with the topology of the Fermi {\it surface}. We demonstrated that the velocities of the collective modes are nonanalytical as topological phase boundaries are crossed. We also identified the locus of spin-charge separation, and showed that, for non-zero spin-orbit coupling and Rabi fields, the collective modes are helical with mixed charge and spin density components. Lastly, we calculated the dynamical structure factors for the charge-charge, spin-charge and spin-spin responses, 
revealing the dispersions, spectral weights 
and helicities of collective modes, paving the way for their experimental detection in systems like $^{6}{\rm Li}$, $^{40}{\rm K}$ and $^{173}{\rm Yb}$.

\end{document}